\documentclass[submission,copyright,creativecommons]{eptcs}
\providecommand{\event}{Refine'15} 
\usepackage{breakurl}             

\usepackage{amssymb}
\usepackage[space]{grffile}
\usepackage{color}
\definecolor{Light}{gray}{.95}
\usepackage[table]{colortbl}
\usepackage{amsmath}
\usepackage{yhmath}

\def\halfthinspace{\relax\ifmmode\mskip.5\thinmuskip\relax\else\kern.8888em\fi}

\newcommand {\defin}{\delta}
\newcommand {\probdist}{\theta}

\newcommand {\emptyrelation}{\phi}

\newcommand {\refinedby}{\sqsubseteq}

\newcommand {\tabeq}{\hspace*{0.2in}}

\newcommand {\tabcom}{\hspace*{1.0in}}

\newcommand {\nln}{@{}l@{}}
\newlength{\interligne}
\newcommand {\dpreuve}{\dimen123=\linewidth \dimen124=\linewidth
\advance\dimen123 by -20mm \advance\dimen124 by -5mm
\advance\dimen123 by -\mathindent \advance\dimen124 by -\mathindent
\setlength{\interligne}{\baselineskip}
\setlength{\baselineskip}{1.2\baselineskip}
    \begin{tabbing} 
    \hspace*{\mathindent}\= \hspace*{5mm}\= \kill 
    \+ \kill}
\newcommand {\fpreuve}{\end{tabbing}
    \setlength{\baselineskip}{\interligne}}

\newcommand {\dpreuveitem}{\begin{tabbing} 
    \hspace*{5mm}\= \hspace*{15mm}\= \kill}
\newcommand {\fpreuveitem}{\end{tabbing}}

\newcommand {\dspecitem}{\begin{tabbing} 
    \hspace*{5mm}\=\hspace*{5mm}\=\hspace*{5mm}\=
    \hspace*{5mm}\=\hspace*{5mm} \kill}
\newcommand {\fspecitem}{\end{tabbing}}

\def\[{\relax\ifmmode\@badmath\else\begin{trivlist}\item[]\leavevmode
  \hbox to\linewidth\bgroup$ \displaystyle
  \hskip\mathindent\bgroup\fi}
\def\]{\relax\ifmmode \egroup $\hfil \egroup \end{trivlist}\else \@badmath \fi}

\newtheorem {proposition}{Proposition}
\newtheorem {definition}{Definition}

\newcommand {\refines}{\sqsupseteq}

\newcommand {\dom}{\mbox{\it dom}}

\newcommand {\dtabin}{\begin{tabbing} 
    \hspace*{\mathindent}\= \kill \+ \kill}
\newcommand {\ftabin}{\end{tabbing}}
\newcommand {\dspec}{\begin{tabbing} 
    \hspace*{\mathindent}\= \hspace*{5mm}\=\hspace*{5mm}\=\hspace*{5mm}\=
                            \hspace*{5mm}\=\hspace*{5mm} \kill 
    \+ \kill}

\newcommand {\fspec}{\end{tabbing}}

\newlength {\longueurtop}
\newcommand {\initlongueurtop}{\setlength{\longueurtop}{\topsep}}
\newcommand {\topzero}{\setlength{\topsep}{0pt}}
\newcommand {\topdefaut}{\setlength{\topsep}{\longueurtop}}
\newcommand {\debuttab}{ \initlongueurtop \topzero \begin{tabbing} }
\newcommand {\fintab}{ \end{tabbing} \topdefaut }

\title{Program Derivation by Correctness Enhancements}
\author{Nafi Diallo
	\institute{New Jersey Institute of Technology,\\
		 Newark, NJ, USA}
	\email{ncd8@njit.edu}
	\and
	Wided Ghardallou
	\institute{University of Tunis El Manar,\\
		 Tunis, Tunisia}
	\email{wided.ghardallou@gmail.com}
	\and
	Jules Desharnais
	\institute{Laval University, Quebec City,\\
		 Quebec, Canada}
	\email{jules.desharnais@ift.ulaval.ca}
	\and
	Ali Mili
	\institute{New Jersey Institute of Technology, \\
		Newark, NJ, USA}
	\email{ali.mili@njit.edu}
}
\def\titlerunning{Program Derivation by Correctness Enhancements}
\def\authorrunning{N. Diallo, W. Ghardallou, J. Desharnais \& A. Mili }
\begin{document}
\maketitle
\begin{abstract}
Relative correctness is the property of a program to be more-correct
than another program with respect to a given specification.  Among the
many properties of relative correctness, that which we found most intriguing
is the property that program $P'$ refines program $P$ if and only if $P'$
is more-correct than $P$ with respect to {\em any} specification.  This inspires
us to reconsider program derivation by successive refinements: each step of
this process mandates that we transform a program $P$ into a program $P'$ that
refines $P$, i.e. $P'$ is more-correct than $P$ with respect to {\em any} specification.
This raises the question:  why should we want to make $P'$ more-correct than $P$
with respect to any specification, when we only have to satisfy specification $R$?
In this paper, we discuss a process of program derivation that replaces
traditional sequence of refinement-based correctness-preserving transformations
starting from specification $R$ by a sequence of relative correctness-based
correctness-enhancing transformations starting from {\tt abort}.
\end{abstract}

\subsection*{Keywords}

Absolute correctness, relative correctness, program
refinement,  program derivation, correctness \\preservation,
correctness enhancement.

\section{Introduction}

\subsection{Background}

Relative correctness is the property of a program to be more-correct
than another program with respect to a given specification.  Intuitively,
$P'$ is more-correct than $P$ with respect to $R$ if and only if $P'$ obeys
$R$ more often (i.e. for a larger set of inputs) than $P$, and violates $R$
less egregiously (i.e. mapping inputs to fewer incorrect outputs)
than $P$.  We have found that relative correctness satisfies
many intuitively appealing properties, such as:  
It is reflexive and transitive, it culminates in absolute correctness, 
and it logically implies enhanced reliability.  Most interesting of all, we
have found that a program $P'$ refines a program $P$ if and only if $P'$
is more-correct than $P$ with respect to {\em any} specification.  
This inspires us to reconsider the process of program derivation by
successive refinements from a specification $R$:  whenever we transform
a program $P$ into a more-refined program $P'$, we are actually mandating
that $P'$ be more-correct than $P$ with respect to {\em any} specification.
This raises the question: why should we impose this condition 
with respect to all
specifications when we have only one specification to satisfy?
Acting on this question, we propose to consider an alternative
process, which we characterize by the following premises:
\begin{itemize}
\item {\em Initial Artifact}.  Whereas in traditional program derivation
we start the stepwise transformation with the specification, in our proposed
derivation we start with the trivial program {\tt abort}, which fails with
respect to any non-empty specification.
\item {\em Intermediate Artifacts}.  Whereas in traditional program derivation
intermediate artifacts are partially defined programs, represented by a mixture
of programming constructs and specification constructs, in our proposed
derivation all intermediate artifacts are finished executable programs.
\item {\em Stepwise Validation}.  Whereas in traditional program derivation
a transformation is considered valid if it proceeds by correctness-preserving
refinement, in our proposed derivation a transformation is considered valid
if it transforms a program into a more-correct program with respect to the
specification we are trying to satisfy.  Because refinement is equivalent to
relative correctness with respect to arbitrary specifications, mandating
relative correctness with respect to a single specification appears to be
a weaker requirement than refinement.
\item {\em Termination Condition}.  Whereas in traditional program derivation
the stepwise transformation ends when we have an executable program, in our
proposed derivation the stepwise transformation ends when we obtain a correct
program; alternatively, if obtaining a correct program is too onerous, and
we are satisfied with a sufficiently reliable program (for a 
given reliability requirement), then this process may
end when the current program's reliability reaches or exceeds the required
threshold.  As we pointed out above, relative correctness logically implies
enhanced reliability, hence the sequence of programs generated by our derivation
process feature monotonically increasing reliability.
\end{itemize}
In the following subsection we discuss the motivation for exploring
this alternative approach to program derivation.

\subsection{Motivation}
\label{motivationsection}

The purpose of this section is to discuss some of the advantages that
our proposed derivation process may offer, by comparison with
traditional refinement-based program derivation.  In the absence of
adequate empirical evidence, all we can do is present some analytical
arguments to the effect that our proposed approach offers some
advantages that may complement those of refinement based program
derivation.
Below are some of the arguments for our position:
\begin{itemize}
\item 
{\em Simpler Transformations}.  
The first argument we offer is that relative correctness 
with respect to a specification $R$ is a weaker
requirement than refinement, for the reason we discussed above:
refinement is equivalent to relative correctness with respect to
{\em all} specification.  Hence we are comparing the condition of
relative correctness with respect to a single specification against
the condition of relative correctness with respect to all
specification.  A simple example illustrates this contrast:  We consider
the following specification $R$ and the following candidate programs,
$P$, $P'$ and $P''$, on a space $S$ defined by two natural variables
$x$ and $y$.  
\begin{itemize}
\item $R=\{(s,s')| x'=x+y\}$,
\item  $P$:  {\tt\{while (y!=0) \{x=x+1; y=y-1;\}\}},
\item  $P'$: {\tt \{x=x+y; y=0;\}},
\item  $P''$: {\tt \{x=x+y;\}}.
\end{itemize}
According to the definitions that we present subsequently, program
{P'} refines program $P$, and program $P''$ is 
more-correct (or as correct as) program $P$ with respect to $R$,
but it does not refine $P$.  As we can see, program $P''$
is simpler than program $P'$ because in fact it is subject to a
weaker requirement:  whereas $P''$ is more-correct than $P$
with respect to $R$, program $P'$ is more-correct than $P$
with respect to all specifications.

\item {\em Keeping Options Open}.  When we derive a program by successive
refinements, every refinement decision restricts the latitude of the
designer for subsequent refinement 
steps.  Consider again the simple example above:
Once we have decided to refine specification $R$ by program $P$, 
we have committed to assign zero to variable $y$, even though the
specification does not require us to do so.  By looking at program $P$,
we have no way to tell which part of the functional attributes of $P$ 
are mandated by the specification (adding $x$ and $y$ into $x$) and
which part stems from previous design decisions (placing 0 into $y$).
By contrast, program derivation by correctness enhancement keeps the
specification in the loop throughout the process, hence maintains
the designer's options intact; in practice, this may come at the cost
of additional complexity; further empirical observation is needed
to assess advantages and drawbacks.

\item {\em A Generic Model}.  Refinement based program derivation can only
be applied to derive a correct program from a specification.  But today
software development from scratch represents a small fraction of software
engineering activity; most software engineering person-months nowadays are
spent on software maintenance and software evolution, and much of software
development involves evolving existing applications rather than developing
new applications from scratch.  We argue that the correctness enhancement
derivation process that we propose captures several software engineering
activities:
\begin{itemize}
\item Software development from scratch:  This is the process we have
discussed in the previous section, that starts from {\tt abort} and
culminates in a correct program.
\item Corrective maintenance:  Corrective maintenance consists in starting
from a program $P$ that is incorrect with respect to a specification
$R$ (which it is intended to satisfy) and mapping it onto a program $P'$
that is more-correct than $P$ with respect to $R$.
\item Adaptive maintenance:  
Adaptive maintenance consists in starting from a program $P$ that
was intended to satisfy some specification $R$ and alter it to now
satisfy a different specification $R'$; this can be modeled as simply
making the program more-correct with respect to $R'$ than it is
currently.  
\item Software upgrade:  Given a specification $R$ and a program $P$, and
given a specification $Q$ that represents a feature we want to integrate
into $P$, upgrading $P$ to satisfy $Q$ amounts to altering $P$ to make it
correct with respect to $Q$ while enhancing or preserving its relative
correctness with respect to $R$.
\item Software evolution:  Given a specification $R$, we want to develop a 
program $P'$ that is correct with respect to $R$; but instead of starting
from scratch, we start from a program $P$ that already satisfies many
requirements of $R$, and process $P$ through correctness-enhancing 
transformations.
\item Deriving reliable software:
For most software products, as for products in general, perfect correctness
is not necessary; very often, adequate reliability (depending on the level of
criticality of the application) is sufficient.  In the program derivation
process by correctness enhancement, deriving a reliable program follows the
same process as deriving a correct program, except that it terminates earlier,
i.e. as soon as the required reliability threshold is matched or exceeded.
\end{itemize}

\item 
{\em Fault Tolerant Derivation Process}.
By design, each transformation in the proposed approach transforms an
intermediate program into a more-correct program; hence if one step
of this process introduces a fault, subsequent steps may well
correct it, since each transformation aims to enhance correctness;
in fact, every subsequent step is an opportunity to correct the
program.
By contrast, in a refinement based (correctness-preserving)
process, a fault in a stepwise transformation effectively dooms
the derivation since all subsequent steps refine a faulty
specification.

\item 
{\em Usable Intermediate Artifacts}.  
In a refinement-based process, only the final artifact is a usable
/ executable program; hence if the process is terminated before its
ultimate step, one has nothing to show for one's effort.  By contrast,
the proposed approach produces a succession of increasingly correct
(hence increasingly reliable) executable
programs.

\end{itemize}
In the remainder of this paper, we briefly introduce the concept of
relative correctness, use it to describe a software development process,
then illustrate it with a simple example.  But first, we need to introduce
some mathematical notations that we use throughout the paper; this is
the subject of the next section.

\section{Mathematical Background}
\label{relationalsect}

\subsection{Relational Notations}

In this section, we introduce some elements of relational mathematics that we use 
in the remainder of the paper to support our discussions; our main source
for definitions and notations is \cite{brinkkahlschmidt97}.  Dealing with programs, 
we represent sets using a programming-like notation, by introducing variable names 
and associated data type (sets of values).  For example, if we represent set $S$ by 
the variable declarations\\
\tabeq $
	x: X; y: Y; z: Z,$\\
then $S$ is the Cartesian product $X\times Y\times Z$.  
Elements of $S$ are denoted in lower case $s$, and are triplets of elements of  $X$, 
$Y$, and $Z$.  Given an element $s$ of $S$, we represent its $X$-component by $x(s)$, 
its $Y$-component by $y(s)$, and its $Z$-component by $z(s)$.  
When no risk of ambiguity exists, we may write $x$ to represent $x(s)$, and $x'$ to
represent $x(s')$, letting the references to $s$ and $s'$ be implicit.

A (binary) relation on $S$ 
is a subset of the Cartesian product $S\times S$; given a pair $(s,s')$ in $R$, 
we say that $s'$ is an {\em image} of $s$ by $R$.  Special relations on $S$ 
include the {\em universal} relation $L=S\times S$, the {\em identity} relation 
$I=\{(s,s')| s'=s\}$, and the {\em empty} relation $\emptyrelation=\{\}$.
Operations on relations (say, $R$ and $R'$) include the set theoretic operations of 
{\em union} ($R\cup R'$), {\em intersection} ($R\cap R'$),
{\em difference} ($R\setminus R'$) and {\em complement}
($\overline{R}$).  They also include the {\em relational product}, denoted by ($
R\circ R'$), or ($RR'$, for short) and defined by:  
$$RR'= \{(s,s')| \exists s'': (s,s'')\in R\wedge (s'',s')\in R'\}.$$
The {\em power} of relation $R$ is denoted by $R^n$, for a natural
number $n$, and defined by $R^0=I$, and for $n>0$, $R^n=R\circ R^{n-1}$.  The
{\em reflexive transitive closure} of relation $R$ is denoted by $R^*$ and defined
by $R^*=\{(s,s')|\exists n\geq 0: (s,s')\in R^n\}$.
The {\em converse} of relation $R$ is the relation denoted by $\widehat{R}$ and 
defined by 
$\widehat{R}=\{(s,s')| (s',s)\in R\}.$
Finally, the {\em domain} of a relation $R$ is defined as the set $\dom(R)=
\{s| \exists s': (s,s')\in R\}$, and the {\em range} of relation $R$ is 
defined as the domain of $\widehat{R}$.

A {\em vector} $R$ is a relation that satisfies the condition $RL=R$; vectors
on set $S$ have the form $A\times S$ for some subset $A$ of $S$.  We use them
as convenient relational representations of sets; for example, given a relation
$R$, the term $RL$ is a vector, which represents the domain of relation $R$.
A {\em monotype} $R$ is a relation that satisfies the condition $R\subseteq I$;
monotypes have the form $\{(s,s')| s'=s\wedge s\in A\}$ for some subset $A$ of $S$;
we represent them by $I(A)$, or by $I(a)$, where $a$ is the characteristic
predicate of set $A$.

\subsection{Refinement Ordering}

The concept of refinement is at the heart of any programming 
calculus; the exact definition of refinement (the property of
a specification to refine another) varies from one calculus
to another; the following definition captures our concept
of refinement.
\begin{definition}
We let $R$ and $R'$ be two relations on space $S$.  We say that
$R'$ {\em refines} $R$ if and only if 
$$RL\cap R'L\cap (R\cup R')=R.$$
\end{definition}
We write this relation as:  $R'\refines R$
or $R\refinedby R'$. Intuitively, $R'$ refines $R$ if and only if $R'$ 
has a larger domain than $R$ and is more deterministic than $R$ inside 
the domain of $R$.
As an illustration of this definition, we let $S$ be the space
defined by $S=\{0,1,2,3\}$ and we let $R$ and $R'$ be the following 
relations:\\
\tabeq $R=\{(1,0),(1,1),(1,2),(2,1),(2,2),(2,3)\}$,\\
\tabeq $R'=\{(0,0),(0,1),(1,1),(1,2),(2,2),(2,1),(3,3),(3,2)\}$.\\
We find:\\
\tabeq $RL=\{1,2\}\times S,$\\
\tabeq $R'L=\{0,1,2,3\}\times S,$\\
whence\\
\tabeq $RL\cap R'L\cap (R\cup R')$\\
= \tabcom \{by inspection, we see $RL\subseteq R'L$\}\\
\tabeq $RL\cap (R\cup R')$\\
= \tabcom \{distributivity\}\\
\tabeq $RL\cap R\cup RL\cap R'$\\
= \tabcom \{since $R\subseteq RL$\}\\
\tabeq $R\cup RL\cap R'$\\
= \tabcom \{by inspection, we see $RL\cap R'\subseteq R$\}\\
\tabeq $R$.\\


\section{Absolute Correctness and Relative Correctness}
\label{absoluterelativecorrectnesssect}
\label{relativecorrectnesssect}

Whereas absolute correctness characterizes the relationship between
a specification and a candidate program,
relative correctness ranks two programs with respect to a
specification; in order to discuss the latter, it helps to
review the former, to see how it is defined in our notation.

\subsection{Program Functions}
\label{programfunctionsect}

Given a program $p$ on space $S$, we denote by $[p]$ the function that $p$
defines on its space, i.e.\\ \tabeq
$P=\{(s,s')| $if program $p$ executes on state $s$ then it terminates
in state $s'\}.$\\
We represent program spaces by means of C-like variable declarations and
we represent programs by means of a few simple C-like programming constructs,
which we present below along with their semantic definitions:
\begin{itemize}
\item {\em Abort:}  $[abort] \equiv  \emptyrelation$.
\item {\em Skip:} $[skip]\equiv I$.
\item {\em Assignment:} $[s=E(s)]\equiv \{(s,s')| s\in\defin(E)\wedge s'=E(s)\}$, where
$\defin(E)$ is the set of states for which expression $E$ can be evaluated.
\item {\em Sequence:} $[p_1; p_2]\equiv [p_1]\circ [p_2]$.
\item {\em Conditional:} $[if~(t)~\{p\}] \equiv T\cap [p]\cup \overline{T}\cap I$, where
$T$ is the vector defined as:  $T=\{(s,s')| t(s)\}$.
\item {\em Alternation:} $[if~(t)~\{p\}~else~\{q\}] \equiv
T\cap [p]\cup \overline{T}\cap[q]$, where $T$ is defined as above.
\item {\em Iteration:} $[while~(t)~\{b\}]\equiv (T\cap [b])^*\cap\widehat{\overline{T}}$,
where $T$ is defined as above.
\item {\em Block:} $[\{x: X; p\}]  \equiv \{(s,s')| \exists x, x': (\langle s,x\rangle,
\langle s',x'\rangle)\in [p]\}$.
\end{itemize}
Rather than use the notation $[p]$ to denote the function of program $p$,
we will usually use upper case $P$ as a shorthand for $[p]$.
By abuse of notation, we may, when it is convenient and causes no
confusion, refer interchangeably to a program and its function
(and we denote both by an upper case letter).

\subsection{Absolute Correctness}

\begin{definition}
\label{correctnessdef}
Let $p$ be a program on space $S$ and let $R$ be a specification on $S$.  
\begin{itemize}
\item We
say that program $p$ is {\em correct} with respect to $R$ if and only if
$P$ refines $R$.
\item We say that program $p$ is {\em partially correct} with respect
to specification $R$ if and only if $P$ refines $R\cap PL$.
\end{itemize}
\end{definition}
This definition is consistent with traditional definitions of partial and total correctness  \cite{hoare69,manna74,gries81,dijkstra76,hehner92}.  
Whenever we want to contrast correctness with partial correctness, we may
refer to it as {\em total correctness}.
The following proposition,
due to \cite{mills86}, gives a simple characterization of correctness,
and sets the stage for the definition of relative correctness.
\begin{proposition}
\label{correctnessprop}
\label{domrpprop}
Program $p$ is correct with respect to specification $R$ if and only if
$(P\cap R)L=RL$.
\end{proposition}
Note that because $(P\cap R)L\subseteq RL$ is a tautology (that stems
from Boolean algebra), the condition above can be written simply as:
$RL\subseteq (P\cap R)L$; this condition can, in turn
(due to relational algebra), be written
merely as,  $R\subseteq (P\cap R)L$.

\subsection{Relative Correctness:  Deterministic Programs}

\begin{figure*}
\thicklines
\setlength{\unitlength}{0.028in}
\begin{center}
\begin{picture}(170,65)

\put(0,0) {\makebox(0,0){6}}
\put(0,10){\makebox(0,0){5}}
\put(0,20){\makebox(0,0){4}}
\put(0,30){\makebox(0,0){3}}
\put(0,40){\makebox(0,0){2}}
\put(0,50){\makebox(0,0){1}}
\put(0,60){\makebox(0,0){0}}

\put(50,0) {\makebox(0,0){6}}
\put(50,10){\makebox(0,0){5}}
\put(50,20){\makebox(0,0){4}}
\put(50,30){\makebox(0,0){3}}
\put(50,40){\makebox(0,0){2}}
\put(50,50){\makebox(0,0){1}}
\put(50,60){\makebox(0,0){0}}

\put(60,0) {\makebox(0,0){6}}
\put(60,10){\makebox(0,0){5}}
\put(60,20){\makebox(0,0){4}}
\put(60,30){\makebox(0,0){3}}
\put(60,40){\makebox(0,0){2}}
\put(60,50){\makebox(0,0){1}}
\put(60,60){\makebox(0,0){0}}

\put(110,0) {\makebox(0,0){6}}
\put(110,10){\makebox(0,0){5}}
\put(110,20){\makebox(0,0){4}}
\put(110,30){\makebox(0,0){3}}
\put(110,40){\makebox(0,0){2}}
\put(110,50){\makebox(0,0){1}}
\put(110,60){\makebox(0,0){0}}

\put(120,0) {\makebox(0,0){6}}
\put(120,10){\makebox(0,0){5}}
\put(120,20){\makebox(0,0){4}}
\put(120,30){\makebox(0,0){3}}
\put(120,40){\makebox(0,0){2}}
\put(120,50){\makebox(0,0){1}}
\put(120,60){\makebox(0,0){0}}

\put(170,0) {\makebox(0,0){6}}
\put(170,10){\makebox(0,0){5}}
\put(170,20){\makebox(0,0){4}}
\put(170,30){\makebox(0,0){3}}
\put(170,40){\makebox(0,0){2}}
\put(170,50){\makebox(0,0){1}}
\put(170,60){\makebox(0,0){0}}

\put(5,50){\vector(4,1){40}}
\put(5,50){\vector(4,-1){40}}
\put(5,40){\vector(4,1){40}}
\put(5,40){\vector(4,-1){40}}
\put(5,30){\vector(4,1){40}}
\put(5,30){\vector(4,-1){40}}
\put(5,20){\vector(4,1){40}}
\put(5,20){\vector(4,-1){40}}
\put(5,10){\vector(4,1){40}}

\put(125,50){\vector(4,1){40}}
\put(125,40){\vector(4,1){40}}
\put(125,30){\vector(4,1){40}}
\put(125,20){\vector(4,1){40}}
\put(125,10){\vector(4,1){40}}
\put(125,0){\vector(4,1){40}}

\put(65,60){\vector(4,-1){40}}
\put(65,50){\vector(4,-1){40}}
\put(65,40){\vector(4,-1){40}}
\put(65,30){\vector(4,-1){40}}
\put(65,20){\vector(4,-1){40}}
\put(65,10){\vector(4,-1){40}}

\put(25,65){\makebox(0,0){$R$}}
\put(85,65){\makebox(0,0){$P$}}
\put(145,65){\makebox(0,0){$P'$}}

\end{picture}
\end{center}
\caption{\label{morecorrectfig}Enhancing Correctness Without Duplicating
Behavior: $P'\refines_R P$}
\end{figure*}
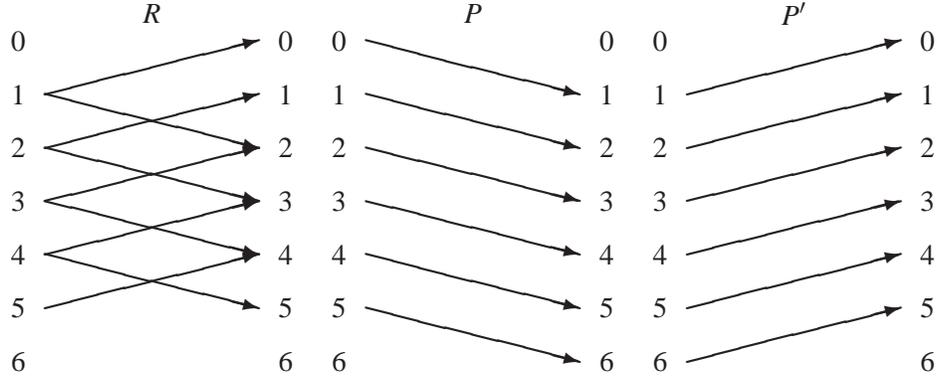

\begin{definition}
Let $R$ be a specification on space $S$ and let $p$ and $p'$ be
two deterministic
programs on space $S$ whose functions are respectively $P$
and $P'$.  
\begin{itemize}
\label{rcdefinition}
\item We say that program $p'$ is {\em more-correct} than program $p$
with respect to specification $R$ (denoted by:  $P'\refines_R P$)
if and only if:
$(R\cap P')L\supseteq (R\cap P)L.$
\item
Also, we say that program $p'$ is {\em strictly more-correct} than
program $p$
with respect to specification $R$ (denoted by:  $P'\sqsupset_R P$) 
if and only if
$(R\cap P')L\supset (R\cap P)L.$
\end{itemize}
\end{definition}
Interpretation:  $(R\cap P)L$ represents (in relational form) the set
of initial states on which the behavior of $P$ satisfies specification
$R$.  We refer to this set as the {\em competence domain} of program $P$.
Relative correctness of $P'$ over $P$ with respect to specification $R$
simply means that $P'$ has a larger competence domain than $P$.
Whenever we want to contrast correctness (given in Definition \ref{correctnessdef})
with relative correctness, we may refer to it as {\em absolute correctness}.
Note that when we say {\em more-correct} we really mean {\em more-correct or
as-correct-as}; we use the shorthand, however, for convenience.
Note that program $p'$ may be more-correct than program $p$ without
duplicating the behavior of $p$ over
the competence domain of $p$: It may have a different behavior (since $R$
is potentially non-deterministic) provided this behavior is also correct with
respect to $R$; see Figure \ref{morecorrectfig}.
In the example
shown in this figure, we have:\\
\tabeq $(R\cap P)L = \{1,2,3,4\}\times S$,\\
\tabeq $(R\cap P')L = \{1,2,3,4,5\}\times S$,\\
where $S=\{0,1,2,3,4,5,6\}$.  
Hence $p'$ is more-correct than $p$ with respect to $R$.

\subsection{Relative Correctness: Non-Deterministic Programs}

The purpose of this section is to define the concept of relative correctness
for arbitrary programs, that are not necessarily deterministic.  One may want
to ask:  why do we need to define relative correctness for non-deterministic
programs if most programming languages of interest are deterministic?  There
are several reasons why we may want to do so:
\begin{itemize}
\item Non-determinacy is a convenient tool to model deterministic programs 
whose detailed
behavior is difficult to capture, unknown, or irrelevant to a particular analysis.
\item We may want to reason about the relative correctness of programs
without having to compute their function is all its minute details.
\item We may want to apply the concept of relative correctness, not only to
finished software products, but also to partially defined intermediate designs
(as appear on a stepwise refinement process).
\end{itemize}
We submit the following definition.
\begin{definition}
\label{nddefinition}
We let $R$ be a specification on set $S$ and we let $P$ and $P'$ be (possibly
non-deterministic) programs on space $S$.  We say that $P'$ is more-correct
than $P$ with respect to $R$ (abbrev:  $P'\refines_R P$) if and only if:
$$(R\cap P)L\subseteq (R\cap P')L\wedge$$
$$(R\cap P)L\cap\overline{R}\cap P'
\subseteq P.$$
\end{definition}

\begin{figure}
\thicklines
\setlength{\unitlength}{0.033in}
\begin{center}
\begin{picture}(140,52)

\put(0,00){\makebox(0,0){5}}
\put(0,10){\makebox(0,0){4}}
\put(0,20){\makebox(0,0){3}}
\put(0,30){\makebox(0,0){2}}
\put(0,40){\makebox(0,0){1}}
\put(0,50){\makebox(0,0){0}}

\put(40,00){\makebox(0,0){5}}
\put(40,10){\makebox(0,0){4}}
\put(40,20){\makebox(0,0){3}}
\put(40,30){\makebox(0,0){2}}
\put(40,40){\makebox(0,0){1}}
\put(40,50){\makebox(0,0){0}}

\put(50,00){\makebox(0,0){5}}
\put(50,10){\makebox(0,0){4}}
\put(50,20){\makebox(0,0){3}}
\put(50,30){\makebox(0,0){2}}
\put(50,40){\makebox(0,0){1}}
\put(50,50){\makebox(0,0){0}}

\put(90,00){\makebox(0,0){5}}
\put(90,10){\makebox(0,0){4}}
\put(90,20){\makebox(0,0){3}}
\put(90,30){\makebox(0,0){2}}
\put(90,40){\makebox(0,0){1}}
\put(90,50){\makebox(0,0){0}}

\put(100,00){\makebox(0,0){5}}
\put(100,10){\makebox(0,0){4}}
\put(100,20){\makebox(0,0){3}}
\put(100,30){\makebox(0,0){2}}
\put(100,40){\makebox(0,0){1}}
\put(100,50){\makebox(0,0){0}}

\put(140,00){\makebox(0,0){5}}
\put(140,10){\makebox(0,0){4}}
\put(140,20){\makebox(0,0){3}}
\put(140,30){\makebox(0,0){2}}
\put(140,40){\makebox(0,0){1}}
\put(140,50){\makebox(0,0){0}}

\put(20,52){\makebox(0,0){$R$}}
\multiput(2,00)(0,10){6}{\vector(1,0){36}}
\multiput(2,00)(0,10){5}{\vector(4,1){36}}
\multiput(2,10)(0,10){5}{\vector(4,-1){36}}

\put(70,52){\makebox(0,0){$P$}}
\multiput(52,00)(0,10){2}{\vector(2,1){36}}
\multiput(52,00)(0,10){2}{\vector(4,3){36}}

\multiput(52,20)(0,10){2}{\vector(4,1){36}}
\multiput(52,20)(0,10){2}{\vector(2,1){36}}

\multiput(52,40)(0,10){2}{\vector(4,-3){36}}
\multiput(52,40)(0,10){2}{\vector(2,-1){36}}

\put(120,52){\makebox(0,0){$P'$}}
\multiput(102,00)(0,10){4}{\vector(2,1){36}}
\put(102,00){\vector(4,3){36}}
\multiput(102,10)(0,10){4}{\vector(4,-1){36}}
\put(102,50){\vector(4,-3){36}}
\put(102,50){\vector(2,-1){36}}
\put(102,40){\vector(2,-1){36}}

\put(50,25){\oval(4,16)}
\put(100,25){\oval(4,36)}

\end{picture}
\end{center}
\caption{\label{rcfigure}Relative Correctness for Non-Deterministic Programs:
$P'\refines_R P$.}
\end{figure}
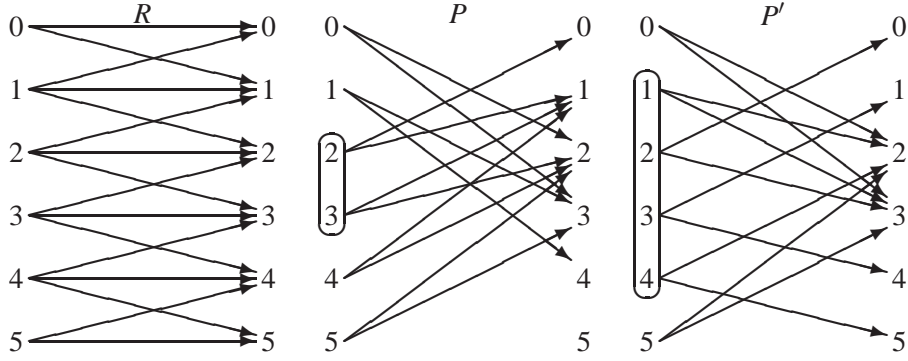

Interpretation:  $P'$ is more-correct than $P$ with respect to $R$ if and only
if it has a larger competence domain, and for the elements in the competence 
domain of $P$, program $P'$ has fewer images that violate $R$ than $P$ does.
As an illustration, we consider the set $S=\{0,1,2,3,4,5,6,7\}$ and we
let $R$, $P$ and $P'$ be defined as follows:\\
\tabeq $R=\{(0,0),(0,1),(1,0),(1,1),(1,2),(2,1),(2,2),(2,3),(3,2),(3,3),(3,4),$\\
\tabeq\tabeq\tabeq  $(4,3),(4,4),(4,5),(5,4),(5,5)\}$\\
\tabeq $P=\{(0,2),(0,3),(1,3),(1,4),(2,0),(2,1),(3,1),(3,2),(4,1),(4,2),(5,2),$\\
\tabeq\tabeq\tabeq $(5,3)\}$\\
\tabeq $P'=\{(0,2),(0,3),(1,2),(1,3),(2,0),(2,3),(3,1),(3,4),(4,2),(4,5),(5,2),$\\
\tabeq\tabeq\tabeq  $(5,3)\}$\\
From these definitions, we compute:\\
\tabeq $R\cap P=\{(2,1),(3,2)\},$\\
\tabeq
$(R\cap P)L=\{2,3\}\times S,$\\
\tabeq $R\cap P'=\{(1,2),(2,3),(3,4),(4,5)\}$\\
\tabeq $(R\cap P')L=\{1,2,3,4\}\times S$\\
\tabeq $(R\cap P)L\cap P'=\{(2,0),(2,3),(3,1),(3,4)\}$\\
\tabeq $(R\cap P)L\cap\overline{R}\cap P'=\{(2,0),(3,1)\}$\\
By inspection, we do find that $(R\cap P)L=\{2,3\}\times S$ is
indeed a subset of  $(R\cap P')L=\{1,2,3,4\}\times S$.  Also, we
find that  $(R\cap P)L\cap\overline{R}\cap P'=\{(2,0),(3,1)\}$
is a subset of $P$.  Hence the 
two clauses of Definition 
\ref{nddefinition} are satisfied.  Figure \ref{rcfigure} represents
relations $R$, $P$ and $P'$ on space $S$.  Program $P'$ is more-correct
than program $P$ with respect to $R$ because it has a larger competence domain
($\{2,3\}$ vs. $\{1,2,3,4\}$, highlighted in Figure
\ref{rcfigure}) and because on the competence
domain of $P$ (=$\{2,3\}$), program $P'$ generates no incorrect output 
($\{(2,0),(3,1)\}$) unless $P$
also generates it.

\section{Program Derivation by Relative Correctness}

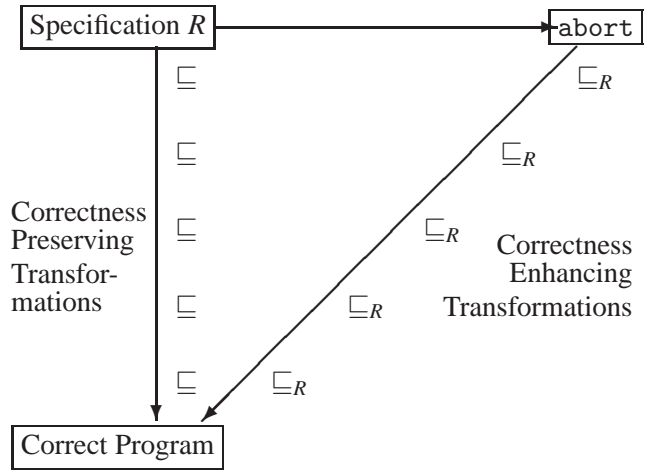
\begin{figure}
\thicklines
\setlength{\unitlength}{0.020in}
\begin{center}
\begin{picture}(120,140)
\put(10,10){\makebox(0,0){\framebox{\shortstack{Correct Program}}}}
\put(10,120){\makebox(0,0){\framebox{\shortstack{Specification $R$}}}}
\put(135,120){\makebox(0,0){\framebox{{\tt abort}}}}

\multiput(25,25)(0,20){5}{$\refinedby$}
\put(0,60){\makebox(0,0){\shortstack[l]{Correctness\\Preserving\\Transfor-\\
mations}}}

\multiput(50,25)(20,20){5}{$\refinedby_R$}
\put(120,55){\makebox(0,0){\shortstack[r]{Correctness\\Enhancing\\
Transformations}}}

\put(130,115){\vector(-1,-1){98}}
\put(35,120){\vector(1,0){90}}
\put(20,115){\vector(0,-1){98}}
\end{picture}
\end{center}
\caption{\label{correctnessenhancingfig}Alternative Program Derivation Paradigms}
\end{figure}

The paradigm of program derivation by relative correctness is
shown in Figure \ref{correctnessenhancingfig}; in this 
section, we illustrate this paradigm on a simple example, where
we show in turn, how to conduct the transformation process
until we find a correct program or (if stakes vs cost considerations
warrant) until we reach a sufficiently reliable program.

\subsection{Producing A Correct Program}

We let space $S$ be defined by three natural variables $n$, $x$ and $y$, and we
let specification $R$ be the following relation on $S$ (borrowed from
\cite{dromey1983}):
$$R = \{(s,s')| n=x'^2-y'^2\wedge 0\leq y'\leq x'\}.$$
Candidate programs must generate $x'$ and $y'$ (if possible) for a given $n$.
The domain of $R$ is the set of states $s$ such that $n(s)$ 
is either odd or a multiple 
of 4; indeed, a multiple of 2 whose half is odd cannot be written as
$n=x'^2-y'^2$, since this equation is equivalent to $n=(x'-y')\times(x'+y')$,
and these two factors ($(x'-y')$ and $(x'+y')$) have the same parity, since
their difference ($x'+y'-x'+y'=2\times y'$) is even.  Hence we write:
$$RL=\{(s,s')| n \mod 2=1\lor n\mod 4=0\}.$$
Starting from the initial program $P_0=${\tt abort}, we resolve to let
the next program $P_1$ be the program that finds this factorization for
$y'=0$:
\begin{verbatim}
void p1()
   {nat n, x, y;   //  input/output variables
    {nat r;         //  work variable
     x=0; y=0; r=0; while (r<n) {r=r+2*x+1; x=x+1;}}}
\end{verbatim}
We compute the function of this program by applying the semantic rules
given in section \ref{programfunctionsect}, and we find:
$$P_1=\{(s,s')| n'=n\wedge y'=0\wedge x'=\lceil\sqrt{n}\rceil\}.$$
Whence we compute the competence domain of $P_1$ with respect to $R$:\\
\tabeq $(R\cap P_1)L$\\
= \tabcom\{substitution, simplification\}\\
\tabeq $\{(s,s')| n=x'^2 \wedge n'=n \wedge y'=0\}\circ L$\\
= \tabcom\{taking the domain\}\\
\tabeq $\{(s,s')| \exists x'':  n=x''^2\}$.\\
In other words, $P_1$ satisfies specification $R$, whenever $n$ is a perfect square.

We now consider the case where $r$ exceeds $n$ by a perfect square, making it
possible to fill the difference with $y^2$; this yields the following
program:
\begin{verbatim}
void p2()
   {nat n, x, y;   //  input/output variables
    {nat r;         //  work variable
    x=0; r=0; while (r<n) {r=r+2*x+1; x=x+1;}
    if (r>n) {y=0; while (r>n) {r=r-2*y-1; y=y+1;}}
    if (r!=n) {abort;}}}
\end{verbatim}
This program preserves $n$, places in $x$ the ceiling of the square root of $n$,
and places in $y$ the integer square root of the difference between $n$ and $x'^2$,
and fails if this square root is not an integer.  We write its function as follows:
$$P_2=\{(s,s')| n'=n\wedge x'=\lceil\sqrt{n}\rceil\wedge y'^2=x'^2-n\wedge y'\geq 0\}.$$
We compute the competence domain of $P_2$ with respect to $R$:\\
\tabeq $(R\cap P_2)\circ L$\\
= \tabcom\{Substitutions\}\\
\tabeq $\{(s,s')| n=x'^2-y'^2\wedge 0\leq y'\leq x'\wedge n'=n\wedge x'=
\lceil\sqrt{n}\rceil \wedge y'^2=x'^2-n$\\
\tabeq\tabeq\tabeq $\wedge y'\geq 0\}\circ L$\\
= \tabcom\{Simplifications\}\\
\tabeq $\{(s,s')| n'=n\wedge x'=\lceil\sqrt{n}\rceil
\wedge y'^2=x'^2-n\wedge y'\geq 0\}\circ L$\\
= \tabcom\{Computing the domain\}\\
\tabeq $\{(s,s')| \exists n'', x'', y'':
n''=n\wedge x''=\lceil\sqrt{n}\rceil\wedge y''^2=x''^2-n\wedge y''\geq 0\}$\\
= \tabcom\{Simplifications\}\\
\tabeq $\{(s,s')| \exists y'':
y''^2=\lceil\sqrt{n}\rceil^2-n\}$.\\
In other words, the competence domain of $P_2$ is the set of states $s$ such that
$n(s)$ satisfies the following property:  the difference between $n(s)$ and the
square of the ceiling of the square root of $n(s)$ is a perfect square.  For example,
a state $s$ such that $n(s)=91$ is in the competence domain of $P_2$, since
$\lceil\sqrt{91}\rceil^2-91=10^2-91=9$, which is a perfect square.
The competence domain of $P_2$ is clearly a superset of the competence domain of 
$P_1$, hence the transition from $P_1$ to $P_2$ is valid.

The next program is derived from $P_2$ by resolving that if the ceiling of
the integer square root of $n$ does not exceed $n$ by a square root, then
we try the next perfect square (whose root we assign to $x$) and we check
whether the difference between that perfect square and $n$ is now a perfect
square; we know that this process converges, for any state $s$ for which
$n(s)$ is odd or a multiple of 4.
This yields the following program:
\begin{verbatim}
void p3()           //  fermat
   {nat n, x, y;   //  input/output variables
    {nat r;         //  work variable
    x=0; r=0; while (r<n) {r=r+2*x+1; x=x+1;}
    while (r>n) 
       {int rsave; y=0; rsave=r;
        while (r>n) {r=r-2*y-1; y=y+1;}
        if (r<n) {r=rsave+2*x+1; x=x+1;}}
   }}
\end{verbatim}
This program preserves $n$, places in $x$ the smallest 
number whose square
exceeds $n$ by a perfect square and places in $y$ the 
square root of the difference between $n$ and $x^2$.  If we let
$\mu(n)$ be the smallest number whose 
square exceeds $n$ by a perfect square, we write the function
of $P_3$ as follows:
$$P_3=\{(s,s')| n'=n\wedge x'=\mu(n)\wedge y'=\sqrt{\mu(n)^2-n}\}.$$
We compute the competence domain of $P$ with respect to $R$:\\
\tabeq $(R\cap P_3)\circ L$\\
= \tabcom\{Substitutions\}\\
\tabeq $\{(s,s')| n=x'^2-y'^2\wedge 0\leq y'\leq x'\wedge n'=n\wedge x'=\mu(n)
\wedge  y'=\sqrt{\mu(n)^2-n}\}\circ L$\\
= \tabcom\{Simplifications\}\\
\tabeq $\{(s,s')| n=x'^2-y'^2\wedge n'=n\wedge x'=\mu(n)\}\circ L$\\
= \tabcom\{Computing the domain\}\\
\tabeq $\{(s,s')| \exists n'', x'', y'': n=x''^2-y''^2\wedge n''=n\wedge x''=\mu(n)\}$\\
= \tabcom\{Simplifications\}\\
\tabeq $\{(s,s')| \exists x'', y'': n=x''^2-y''^2\}$\\
= \tabcom\{By inspection\}\\
\tabeq $RL$.\\
Hence $P_3$ is correct with respect to $R$ (by proposition \ref{correctnessprop})
hence it is more-correct than $P_2$ with respect to $R$.  Hence we do have:
$$P_0\refinedby_R P_1\refinedby_R P_2\refinedby_R P_3.$$
Furthermore, we find that $P_3$ is correct with respect to $R$; this concludes
the derivation.

\subsection{Producing A Reliable Program}

We interpret the reliability of a program as the probability of a
successful execution 
of the program on some initial state selected
at random from the domain of $R$ according to some probability
distribution $\probdist$.  Given a probability distribution
$\probdist$ on $dom(R)$, the reliability of a candidate program
$P$ is then the probability that an element of $dom(R)$ selected
according to the probability distribution $\probdist$ falls in the
competence domain of $P$ with respect to $R$.  Clearly, the larger
the competence domain, the higher the probability.  Hence the 
sequence of programs that we generate in the proposed process
feature higher and higher reliability.  So that if we are supposed to
derive a program under a reliability requirement, we can terminate 
the stepwise transformation process as soon as we obtain a
program whose estimated reliability matches or exceeds the
specified threshold.  So far this is a theoretical proposition,
but an intriguing possibility nevertheless.  The sample
program developed in the previous subsection may be used to
illustrate this idea, though it does not show a uniform reliability
growth.  For the sake of argument, we suppose that $n$ ranges between
1 and 10000, and we estimate the reliability of each of the programs
generated in the transformation process.
\begin{itemize}
\item $P_0$: The reliability of $P_0$ is zero, of course, since it never runs
successfully.
\item $P_1$: If $n$ takes values between 1 and 10000, then the domain of $R$
has 7500 elements (since 1 out of four is excluded: even numbers whose half is
odd are not decomposable); out of these 7500 elements, only 100 are perfect
squares ($1^2$ to $100^2$).  
Hence the reliability of $P_1$ under a uniform probability
distribution is $\frac{100}{7500}=0.01333$.
\item $P_2$: The competence domain of $P_2$ includes all the elements $n$
that can be written as:  $n=\lceil\sqrt{n}\rceil^2-y^2$ for some non-negative
value $y$.  To count the number of such elements, we consider all possible
values of $x$ (between 1 and 100) and all possible values of $y$ such that
$(x-1)^2<x^2-y^2\leq x^2.$  By inverting the inequalities and adding $x^2$ to
all sides, we obtain:
$$0\leq y^2<2x-1.$$
Hence the number of elements in the competence domain of $P_2$ can be written as
$$100 + \sum_{x=1}^{100} \sqrt{2x-1}.$$
We find this quantity to be equal to $996$, which yields a probability of $
\frac{996}{7500}=0.1328$.
\item $P_3$: Because the competence domain of $P_3$ is all of $dom(R)$, the
reliability of this program is 1.0.  
\end{itemize}
We obtain the following table.
\begin{center}
\begin{tabular}{|l|r|}
\hline
\hline
Program & Reliability \\
\hline\hline
$P_0$ & 0.0000 \\
\hline
$P_1$ & 0.0133 \\
\hline
$P_2$ & 0.1328 \\
\hline
$P_3$ & 1.0000 \\
\hline\hline
\end{tabular}
\end{center}

\section{Conclusion}

\subsection{Summary}

In this paper, we argue that program derivation by successive refinements
may, perhaps, be imposing an unnecessarily strong condition on each step
of the transformation process; also, we submit that by using the weaker
criterion of relative correctness rather than refinement we may be achieving
greater flexibility in the design process, and perhaps simpler solutions,
without loss of quality.  
With hindsight, the proposed approach appears to be a natural alternative:
Indeed, if we want to derive a correct program from a given specification,
we can either transform the specification in a correctness-preserving manner
until it becomes a program, or start from a trivial program and transform
it in a correctness-enhancing manner until it becomes correct.
A simple way to contrast these two paradigms is to model them as
iterative processes, and to characterize each one of them by:
its initial state, its invariant assertion, its variant function, and its
exit condition; this is shown below.
\begin{center}
\begin{tabular}{|l|l|l|}
\hline\hline
Attribute & Refinement Based & Relative Correctness Based\\
\hline\hline
Initialization & $a$ = $R$ & $a$ = {\tt abort} \\
\hline
Invariant      & $a$ is correct & $a$ is a program \\
\hline
Variant        & $a$ increasingly concrete & $a$ increasingly correct \\
\hline
Exit test      & when $a$ is a program & when $a$ is correct\\
\hline\hline
\end{tabular}
\end{center}
The proposed paradigm appears to model several software
engineering activities, including:  the development of (sufficiently)
reliable programs; corrective maintenance; adaptive maintenance; software
upgrade; and software evolution.  Hence by advancing the state of the
art in correctness-enhancing program derivation, we stand to have
a greater impact on software engineering practice than if we focus exclusively
on correctness-preserving program derivation.
We have illustrated our thesis by a simple
example, although we admit than this example does not constitute evidence
of viability.  

\subsection{Related Work}

While, to the best of our knowledge, our work is the first to apply 
relative correctness to program derivation, it is not the first to
introduce a concept of relative correctness.
In \cite{logozzo2014} Logozzo discusses a framework for ensuring that
some semantic properties are preserved by program transformation
in the context of software maintenance.
In \cite{lahiri2013} Lahiri et al. 
present a technique for verifying the
relative correctness of a program with respect to a previous
version, where they represent specifications by means of 
executable assertions placed throughout the program, and they
define relative correctness by means of inclusion relations
between sets of successful traces and unsuccessful traces.
Logozzo and Ball \cite{logozzo2012} take a similar approach to
Lahiri et al. in the sense that they represent specifications 
by a network of executable assertions placed throughout the
program, and they define relative correctness in terms of
successful traces and unsuccessful traces of candidate
programs.
Our work differs significantly from all these works in many ways: first,
we use relational specifications that address the functional properties 
of the program as a whole, and are not aware of intermediate 
assertions that are expected to hold throughout the program;  second, our
definition of relative 
correctness involves competence domains (for deterministic
specifications) and the sets of states that candidate programs produce
in violation of the specification (for non-deterministic programs); third
we conduct a detailed analysis of the relations between relative correctness
and the property of refinement.

Also related to our work are proposals by Banach and Pempleton
\cite{banach2000} and by Prabhu et al. \cite{ghosal,deshmukh2015,chatterjee2015}
to find alternatives for strict refinement-based program
derivation.  In \cite{banach2000} Banach and Pempleton introduce the
concept of {\em retrenchment}, which is a property linking two successive
artifacts in a program derivation, that are not necessarily ordered by
refinement; the authors argue that strict refinement may sometimes be 
inflexible, and present retrenchment as a viable substitute, that trades
simplicity for strict correctness preservation, and discuss under what
conditions the substitution is viable.  In \cite{ghosal,deshmukh2015,%
chatterjee2015} Prabhu et al. propose another alternative to strict refinement,
which is {\em approximate refinement}.  Whereas strict refinement defines
a partial ordering between artifacts, whereby a concrete artifact is a
correctness-preserving
implementation for an abstract artifact, approximate refinement
defines a topological distance between artifacts, and considers
that a concrete implementation is acceptable if it is close enough
(by some measure of distance) to the abstract artifact.
Retrenchment and Approximate refinement are both substitutes for refinement
and are both used in a correctness-preserving transformation from a 
specification to a program; by contrast, relative correctness offers
an orthogonal paradigm that seeks correctness enhancement rather than
correctness preservation.

\subsection{Prospects}

In this paper we merely suggested an alternative paradigm for the
derivation of correct (or reliable) programs from a specification; we
neither showed, through empirical evidence, that this is a viable
alternative, nor showed how to apply it in general.  These two questions
are the most pressing issues in our research agenda.

\subsubsection*{Acknowledgments}

The authors gratefully acknowledge the valuable and
insightful feedback provided by the
anonymous reviewers.

\bibliographystyle{eptcs}
\bibliography{derref}

\begin{thebibliography}{10}
\providecommand{\bibitemdeclare}[2]{}
\providecommand{\surnamestart}{}
\providecommand{\surnameend}{}
\providecommand{\urlprefix}{Available at }
\providecommand{\url}[1]{\texttt{#1}}
\providecommand{\href}[2]{\texttt{#2}}
\providecommand{\urlalt}[2]{\href{#1}{#2}}
\providecommand{\doi}[1]{doi:\urlalt{http://dx.doi.org/#1}{#1}}
\providecommand{\bibinfo}[2]{#2}

\bibitemdeclare{inproceedings}{banach2000}
\bibitem{banach2000}
\bibinfo{author}{R.~\surnamestart Banach\surnameend} \&
  \bibinfo{author}{M.~\surnamestart Poppleton\surnameend}
  (\bibinfo{year}{2000}): \emph{\bibinfo{title}{Retrenchment, Refinement and
  Simulation}}.
\newblock In: {\sl \bibinfo{booktitle}{ZB: Formal Specifications and
  Development in Z and B}}, \bibinfo{series}{Lecture Notes in Computer
  Science}, \bibinfo{publisher}{Springer}, pp. \bibinfo{pages}{304--323},
  \doi{10.1007/3-540-44525-0\_18}.

\bibitemdeclare{book}{brinkkahlschmidt97}
\bibitem{brinkkahlschmidt97}
\bibinfo{author}{Ch. \surnamestart Brink\surnameend},
  \bibinfo{author}{W.~\surnamestart Kahl\surnameend} \&
  \bibinfo{author}{G.~\surnamestart Schmidt\surnameend} (\bibinfo{year}{1997}):
  \emph{\bibinfo{title}{Relational Methods in Computer Science}}.
\newblock \bibinfo{publisher}{Springer Verlag},
  \doi{10.1007/978-3-7091-6510-2}.

\bibitemdeclare{article}{chatterjee2015}
\bibitem{chatterjee2015}
\bibinfo{author}{K.~\surnamestart Chaterjee\surnameend} \&
  \bibinfo{author}{Vinayak~S. \surnamestart Prabhu\surnameend}
  (\bibinfo{year}{2015}): \emph{\bibinfo{title}{Quantitative Temporal
  Simulation and Refinement Distancess for Timed Systems}}.
\newblock {\sl \bibinfo{journal}{IEEE Transactions for Automatic Control}}
  \bibinfo{volume}{60}(\bibinfo{number}{9}), pp. \bibinfo{pages}{2291--2306},
  \doi{10.1109/TAC.2015.2404612}.

\bibitemdeclare{inproceedings}{deshmukh2015}
\bibitem{deshmukh2015}
\bibinfo{author}{J.~V. \surnamestart Deshmukh\surnameend},
  \bibinfo{author}{R.~\surnamestart Majumdar\surnameend} \&
  \bibinfo{author}{V.~\surnamestart Prabhu\surnameend} (\bibinfo{year}{2015}):
  \emph{\bibinfo{title}{Quantifying Conformance Using the Skorokhod Metric}}.
\newblock In: {\sl \bibinfo{booktitle}{Proceedings, CAV: Computer Aided
  Verification}}, \bibinfo{publisher}{Springer Verlag}, pp.
  \bibinfo{pages}{234--250}, \doi{10.1007/978-3-319-21668-3\_14}.

\bibitemdeclare{book}{dijkstra76}
\bibitem{dijkstra76}
\bibinfo{author}{E.W. \surnamestart Dijkstra\surnameend}
  (\bibinfo{year}{1976}): \emph{\bibinfo{title}{A Discipline of Programming}}.
\newblock \bibinfo{publisher}{Prentice Hall}.

\bibitemdeclare{techreport}{dromey1983}
\bibitem{dromey1983}
\bibinfo{author}{G.~\surnamestart Dromey\surnameend} (\bibinfo{year}{1983}):
  \emph{\bibinfo{title}{Program Development by Inductive Stepwise Refinement}}.
\newblock \bibinfo{type}{Technical Report} \bibinfo{number}{Working Paper
  83-11}, \bibinfo{institution}{University of Wollongong, Australia},
  \doi{10.1002/spe.4380150102}.

\bibitemdeclare{techreport}{ghosal}
\bibitem{ghosal}
\bibinfo{author}{A.~\surnamestart Ghosal\surnameend},
  \bibinfo{author}{M.~\surnamestart Jurdzinski\surnameend},
  \bibinfo{author}{R.~\surnamestart Majumdar\surnameend} \&
  \bibinfo{author}{Vinayak \surnamestart Prabhu\surnameend}
  (\bibinfo{year}{2005}): \emph{\bibinfo{title}{Approximate Refinement for
  Hybrid Systems}}.
\newblock \bibinfo{type}{Technical Report}, \bibinfo{institution}{University of
  California at Berkeley}.

\bibitemdeclare{book}{gries81}
\bibitem{gries81}
\bibinfo{author}{D.~\surnamestart Gries\surnameend} (\bibinfo{year}{1981}):
  \emph{\bibinfo{title}{The Science of programming}}.
\newblock \bibinfo{publisher}{Springer Verlag},
  \doi{10.1007/978-1-4612-5983-1}.

\bibitemdeclare{book}{hehner92}
\bibitem{hehner92}
\bibinfo{author}{E.C.R. \surnamestart Hehner\surnameend}
  (\bibinfo{year}{1992}): \emph{\bibinfo{title}{A Practical Theory of
  Programming}}.
\newblock \bibinfo{publisher}{Prentice Hall}, \doi{10.1007/978-1-4419-8596-5}.

\bibitemdeclare{article}{hoare69}
\bibitem{hoare69}
\bibinfo{author}{C.A.R. \surnamestart Hoare\surnameend} (\bibinfo{year}{1969}):
  \emph{\bibinfo{title}{An axiomatic basis for Computer programming}}.
\newblock {\sl \bibinfo{journal}{Communications of the ACM}}
  \bibinfo{volume}{12}(\bibinfo{number}{10}), pp. \bibinfo{pages}{576 -- 583},
  \doi{10.1145/363235.363259}.

\bibitemdeclare{inproceedings}{lahiri2013}
\bibitem{lahiri2013}
\bibinfo{author}{S.~K. \surnamestart Lahiri\surnameend}, \bibinfo{author}{K.~L.
  \surnamestart McMillan\surnameend}, \bibinfo{author}{R.~\surnamestart
  Sharma\surnameend} \& \bibinfo{author}{C.~\surnamestart
  Hawblitzel\surnameend} (\bibinfo{year}{2013}):
  \emph{\bibinfo{title}{Differential Assertion Checking}}.
\newblock In: {\sl \bibinfo{booktitle}{Proceedings, ESEC/ SIGSOFT FSE}}, pp.
  \bibinfo{pages}{345--455}, \doi{10.1145/2491411.2491452}.

\bibitemdeclare{inproceedings}{logozzo2012}
\bibitem{logozzo2012}
\bibinfo{author}{F.~\surnamestart Logozzo\surnameend} \&
  \bibinfo{author}{T.~\surnamestart Ball\surnameend} (\bibinfo{year}{2012}):
  \emph{\bibinfo{title}{Modular and Verified Automatic Program Repair}}.
\newblock In: {\sl \bibinfo{booktitle}{Proceedings, OOPSLA}}, pp.
  \bibinfo{pages}{133--146}, \doi{10.1145/2384616.2384626}.

\bibitemdeclare{inproceedings}{logozzo2014}
\bibitem{logozzo2014}
\bibinfo{author}{F.~\surnamestart Logozzo\surnameend},
  \bibinfo{author}{S.~\surnamestart Lahiri\surnameend},
  \bibinfo{author}{M.~\surnamestart Faehndrich\surnameend} \&
  \bibinfo{author}{S.~\surnamestart Blackshear\surnameend}
  (\bibinfo{year}{2014}): \emph{\bibinfo{title}{Verification Modulo Versions:
  Towards Usable Verification}}.
\newblock In: {\sl \bibinfo{booktitle}{Proceedings, PLDI}},
  p.~\bibinfo{pages}{32}, \doi{10.1145/2594291.2594326}.

\bibitemdeclare{book}{manna74}
\bibitem{manna74}
\bibinfo{author}{Z.~\surnamestart Manna\surnameend} (\bibinfo{year}{1974}):
  \emph{\bibinfo{title}{A Mathematical Theory of Computation}}.
\newblock \bibinfo{publisher}{McGraw Hill}.

\bibitemdeclare{book}{mills86}
\bibitem{mills86}
\bibinfo{author}{H.D. \surnamestart Mills\surnameend}, \bibinfo{author}{V.R.
  \surnamestart Basili\surnameend}, \bibinfo{author}{J.D. \surnamestart
  Gannon\surnameend} \& \bibinfo{author}{D.R. \surnamestart Hamlet\surnameend}
  (\bibinfo{year}{1986}): \emph{\bibinfo{title}{Structured Programming: A
  Mathematical Approach}}.
\newblock \bibinfo{publisher}{Allyn and Bacon}, \bibinfo{address}{Boston, Ma}.

\end{thebibliography}
\end{document}